\documentclass[runningheads,a4paper]{llncs}

\usepackage{amssymb}
\usepackage{graphicx}

\usepackage{url}
\urldef{\mailsa}\path|sani9585@gmail.com |
\urldef{\mailsb}\path|bouk@comsats.edu.pk |
\urldef{\mailsc}\path|amjadiitkust@gmail.com|
\urldef{\mailsd}\path|nadeemjavaid@comsats.edu.pk|
\urldef{\mailse}\path|sasase@ics.keio.ac.jp | 
   
\newcommand{\keywords}[1]{\par\addvspace\baselineskip
\noindent\keywordname\enspace\ignorespaces#1}
\begin{document}
\mainmatter  
\title{Effect of Fast Moving Object on RSSI in WSN:\\An Experimental Approach}
\titlerunning{Effect of Fast Moving Object on RSSI in WSN:\textit{An Experimental Approach}}
%
%
\author{Syed Hassan Ahmed $^{\dag a}$%
\and Safdar H. Bouk $^{\ddag b}$ \and Amjad Mehmood $^{\dag c}$ \and Nadeem Javaid $^{\ddag d}$ \and Sasase Iwao $^{\S e}$}
\authorrunning{Effect of Fast Moving Object on RSSI in WSN:\textit{An Experimental Approach}}

\institute{$^{\dag}$Institute of IT, Kohat Univ. of Science and Technology,Kohat, Pakistan.\\
$^{\ddag}$Dept. of Electrical Engineering COMSATS Institute of Information Technology, Islamabad, Pakistan.\\$^{\S}$Department of Information and Computer Science,
Keio University, Japan.\\
$^{a}$\mailsa  \ $ ^{b}$\mailsb  \ $ ^{c}$\mailsc \ $ ^{d}$\mailsd \ $ ^{e}$\mailse}

%
%

\toctitle{Lecture Notes in Computer Science}
\tocauthor{Authors' Instructions}
\maketitle

\begin{abstract}
In this paper, we experimentally investigate the effect of fast moving object on the RSSI in the wireless sensor networks in presence of the ground effect and antenna orientation in elevation direction. In experimental setup, MICAz mote pair was placed on the ground, where one mote acts as a transmitter and the other as a receiver. The transmitter mote's antenna was oriented in elevation direction with respect to the receiver mote's antenna. The fast moving object i.e. car, was passed between the motes and the fluctuations in the RSSI are observed. The experimental results show some sequential pattern in RSSI fluctuations when car moves at some relatively slow speed. However, some irregularities were also observed when antenna was oriented at 45  and 90  in elevation direction.
\keywords{RSSI; MICAz; WSN; Sensor-less WSN}
\end{abstract}

\section{Introduction}

It has been observed that Wireless Sensor Networks (WSN) is one of the dynamic research area because, it has been investigated by many researchers due to its diversified applications i.e. health, traffic, agriculture, military, goods tracking, etc. WSN comprises low-power wireless nodes or motes and gateway node(s) \cite{01}. The task of wireless node is to sense the environment or its physical parameters (that is done with the help of sensors or sensing modules attached with the sensor nodes) and communicate the sensed data to the gateway node(s). Gateway node(s) can either be connected to the computer or directly to the network. The sensed data is either stored in a central online database or at the computer that is attached with the gateway node(s).

There are several areas that are intensively investigated by the researchers, which includes sensor platform design, energy efficient routing and MAC protocols, deployment strategies and the list goes on. Along with these applications, an active research is also being carried out where sensing is accomplished without using any physical sensors or sensing modules and termed as Sensor-less WSN (SL-WSN) \cite{02}.

In SL-WSN, no sensing module is attached to the mote and sensing is performed by monitoring and measuring the variations of the power in received radio signal at the receiving antenna. This signal strength metric is called Receive Signal Strength Indicator (RSSI). There are different factors that affect the RSSI value, which includes antenna type, physical distance between sender and receiver, physical objects, interference from other RF devices, transmit power of sender mote, antenna orientation and so on.

In this paper, we experimentally investigate the exact impact of fast moving object i.e. vehicle, on the RSSI of the mote pair. During the experiment, we use MICAz (MPR2600J Japan) motes \cite{3} and MIB520 gateway and programming board. Results are analyzed on the basis of different vehicle speeds and elevation angles of the MICAz mote antenna. The ground effect is also considered in the experiment by placing the MICAz mote pair on the ground.

Following section, briefly discusses the related work. Section-\ref{sec:expsetup}, describes our experimental approach to monitor instabilities in RF signals using Crossbow Micaz motes and realistic data collection. Experimental results are discussed in Section-\ref{sec:results}. Finally, Section-\ref{sec:conclusion} concludes the paper.

\section{Related Work}\label{sec:relwork}
RSSI of MICAz mote has been investigated several times in the past, where the effects of human activities, distance and antenna position, on RF propagations have been observed. Also, the RSSI has been considered as one of the important factor to compute the radio link quality in the wireless networks. Several routing and localization protocols rely severely on the link quality metric, for example, to choose the best wireless link during the routing process or to estimate the distance between two communicating sensors. 
Srinivasan et al. in \cite{4} considered MICAz motes and showed that the RSSI values can be a promising connection quality indicator if its value is above the predefined sensitivity threshold of the radio frequency module. In case, if RSSI value is near this threshold then, it does not have a correlation with packet reception rate and it is only because of the local noise variations at nodes.
The TinyOS based radio performance of WSN is investigated in \cite{5}. It is shown that the antenna of the Tmote Sky device is not perfectly omni-directional and the RSSI value above a certain point should not be used as a distance indicator between sensors. 
Pius W.Q. Lee et. al. \cite{6} considers variations in the received signal strength or RSSI value caused by the passive objects, e.g. humans, to monitor and detect the human activity in the network deployment area. RSSI value is analyzed in different experimental scenarios e.g. indoor, large hall, outdoor etc. During these experiments nodes were placed at 3m to 5m apart and 1.5m above the ground and no antenna orientation has been considered in the experiments.
The impact of human activity is also investigated experimentally through different experimental scenarios in \cite{7}. It is also observed that the RSSI is not the robust link quality indicator and human activity has great impact on the packet reception rate. 
In \cite{8}, authors investigated the impact of antenna orientation on the performance of WSN. The antenna of both the MICAz mote pair is oriented in elevation direction at different angles to the ground and their impact on the RSSI is estimated.
The previous work discussed above, doesn't consider the effect of the fast moving object on the RSSI in presence of the ground effect and the antenna orientation in elevation direction. In this paper, we experimentally investigate the effect of vehicle on the RSSI when nodes are placed on the ground and the antenna of the transmitting node is elevated in the elevation direction.

\begin{figure}
\center
\includegraphics[height=4.2cm]{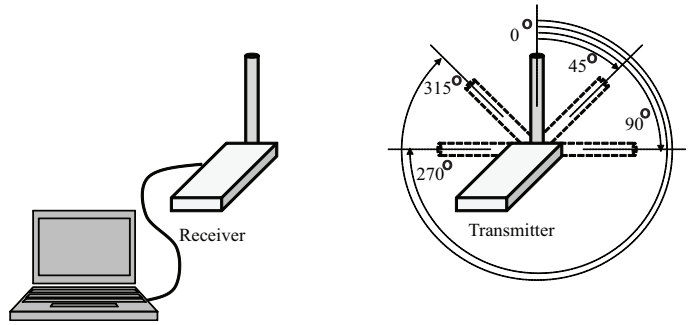}
\caption{Antenna orientation of the transmitter node.}
\label{Fig_2}
\end{figure} 

\section{Experimental Setup}\label{sec:expsetup}

This section briefly describes the experimental setup. In our experiment, we use MICAz OEM Edition motes (MPR2600J Japan), by MEMSIC Inc., formerly Crossbow. The MICAz OEM Edition mote is equipped with Atmega128L processor, CC2420 RF Chip and a half-wave external monopole antenna. CC2420 is designed for the low-power and low-rate systems that operate at ISM Frequency Band from 2400MHz to 2483.5MHz.

\begin{figure}
\center
\includegraphics[height=8.2cm]{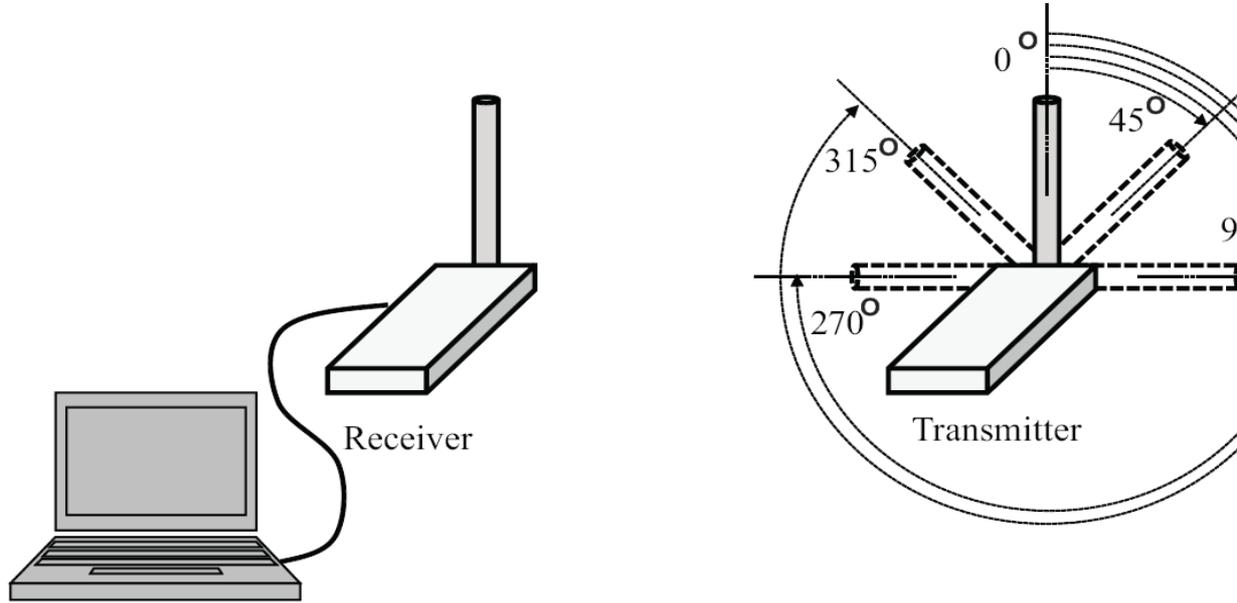}
\caption{Experimental setup where Car is passing between MICAz Mote pair}
\label{Fig_1a}
\end{figure} 

During the experiment we analyze the effect of fast moving object i.e. Car, over the RSSI in presence of ground effect and antenna orientation. One of the motes works as a transmitter mote (that continuously sends packets at the fixed interval of 50ms) and the other node acts as a receiver. These nodes are placed on the ground at 5m apart. We drive the car in between LOS of these nodes with varying speed of 10, 20, to 60km/h. Also antenna of the transmitter node is orientated at 0, 45, 90 to 315° in elevation direction with respect to the receiver mote.
 The RSSI of the successfully received packets at the receiver mote is recorded in a computer attach with it. Fig. 1 and Fig. 2 show the antenna orientation and experimental setup where car is passing through MICAz mote pair.

\section{Experimental Results}\label{sec:results}

\begin{figure}
\center
\includegraphics[height=4.2cm]{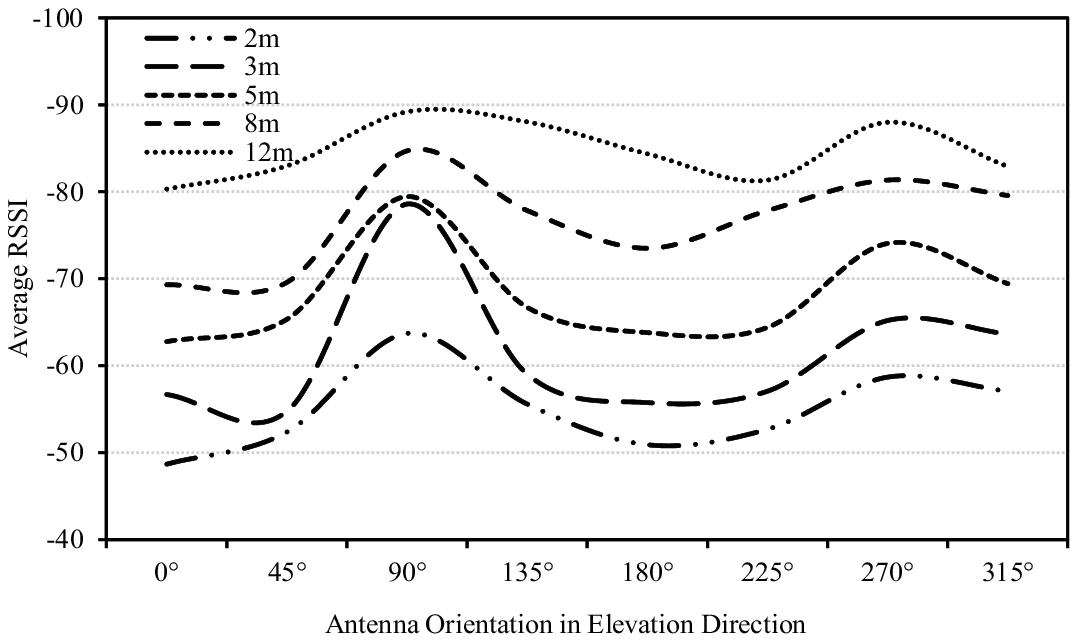}
\caption{Average RSSI vs. antenna orientation in elevation direction without any object movement}
\label{fig_3}
\end{figure} 

\begin{figure}
\center
\includegraphics[height=4.2cm]{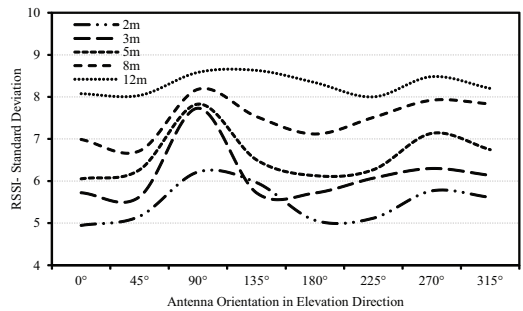}
\caption{RSSI standard deviation vs. antenna orientation in elevation direction without any object movement}
\label{Fig_4}
\end{figure}

In this section, experimental results are briefly discussed. Initially, during the experiment, the average RSSI between node pair (transmitter and receiver), placed at varying distances and with antenna orientation at different angles in elevation direction, has been observed with no vehicle movement and are shown in Fig. 3 and Fig. 4. Fig. 3 and 4 show the average RSSI and Standard Deviation in RSSI vs. antenna orientation, respectively. It is evident from the results that RSSI decreases and deviation in RSSI increases as distance increases. However, some major RSSI fluctuations at 90° and 270° have been observed in all RSSI readings.

The obvious reason of those rise and dips is that monopole antenna radiates power in perpendicular direction to the antenna itself. When both antennas are placed in parallel position to each other, the radiation power increases. On the other hand, when one antenna is rotated in elevation direction with reference to the other, radiation power decreases at the receiver side. In result, the RSSI decreases when antenna is oriented at 90 and 270 elevation direction.

\begin{figure}
\center
\includegraphics[height=8.2cm]{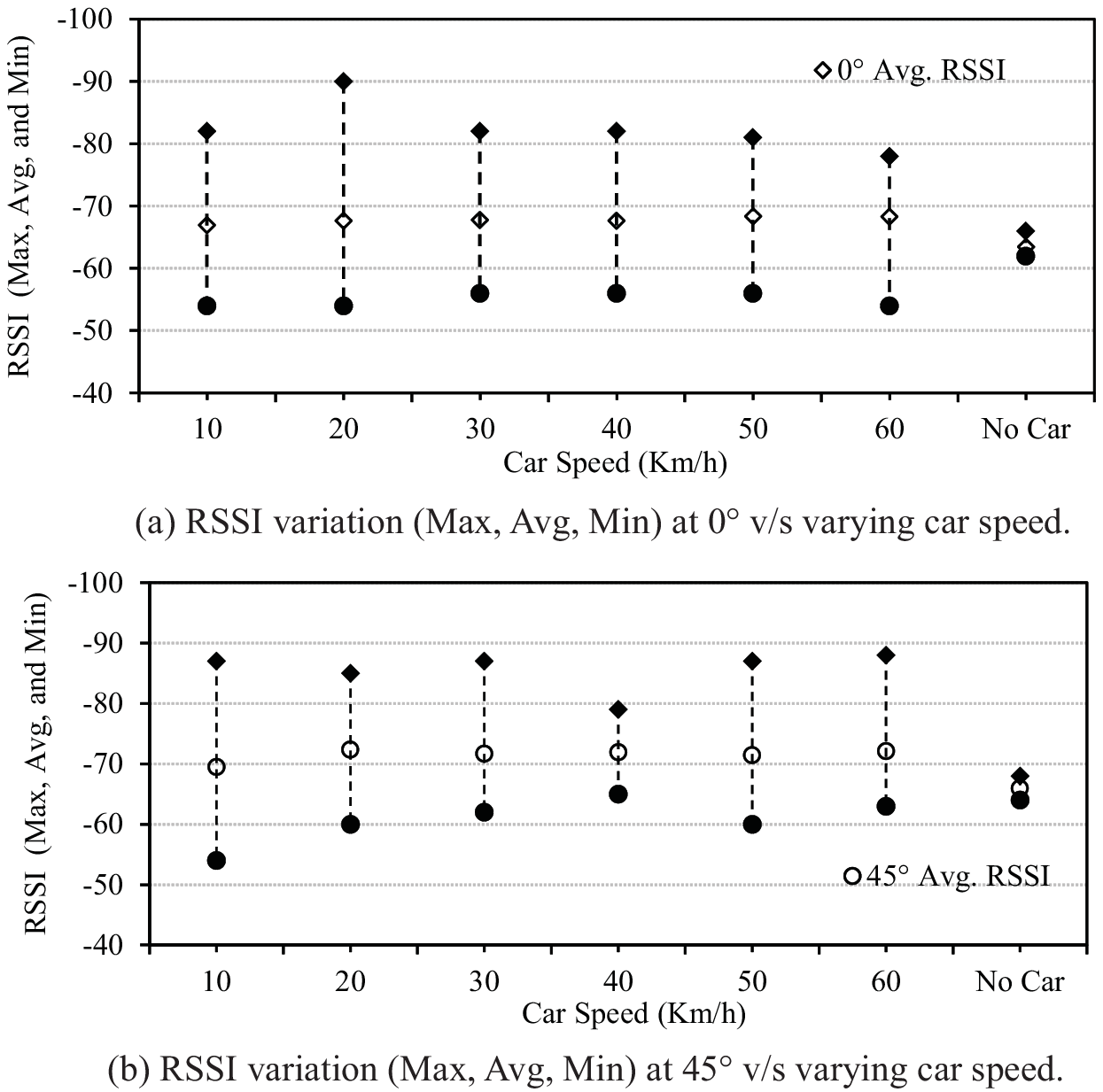}
\caption{RSSI variation vs. varying car speed at (a) 0  orientation and (b) 45  orientation, in elevation direction.}
\label{Fig_5}
\end{figure}

\begin{figure}
\center
\includegraphics[height=8.2cm]{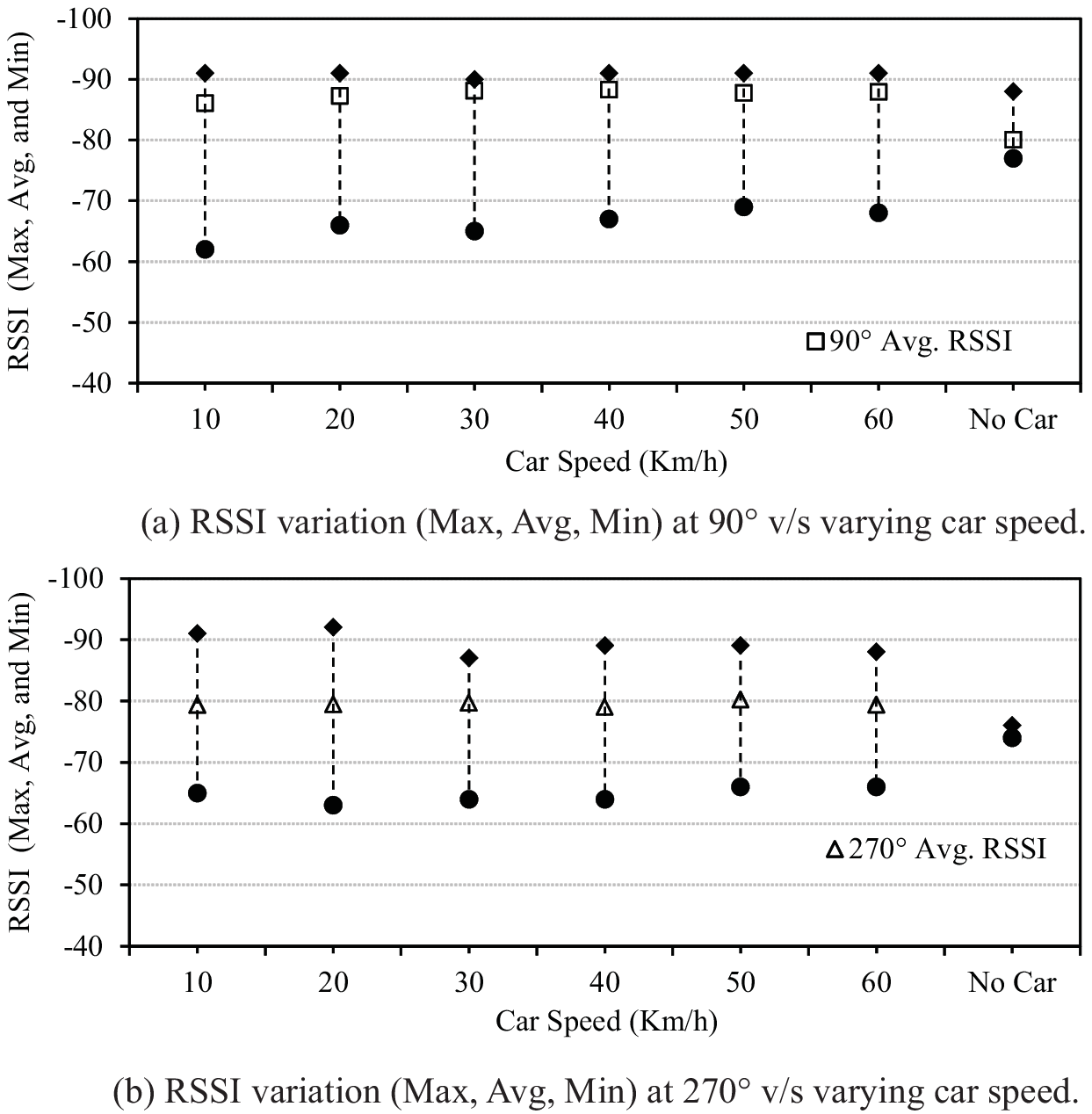}
\caption{RSSI variation vs. varying car speed at (a) 90  orientation and (b) 270  orientation, in elevation direction}
\label{Fig_6}
\end{figure}

The effect of the vehicle (car) moving at varying speed between the motes pair placed on the ground at 5m apart is observed during the experiment. The maximum fluctuation, average and minimum fluctuation in RSSI values with varying car speed at 0 , 45  and 90 , 270  are shown in Fig. 5 and Fig. 6, respectively.  It is evident from the results that when car is moving at slow speed e.g. 10 and 20km/h the distance between RSSI boundary values (Max. and Min.) is larger.

\begin{figure}
\center
\includegraphics[height=4.2cm]{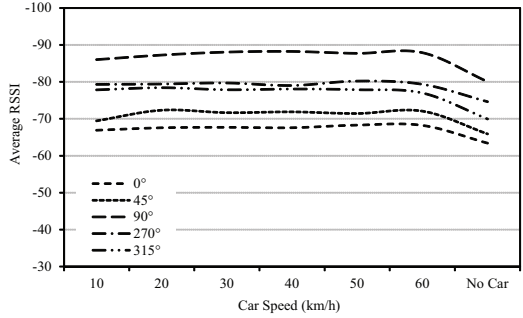}
\caption{Average RSSI vs. varying car speed for different antenna orientation angles.}
\label{Fig_7}
\end{figure}

However, if the car speed is faster than 20km/h then, the boundary values are almost similar or sometimes irregular (e.g. 40km/h speed at 45  antenna orientation in Fig. 5(b)). However, these fluctuation in RSSI when car moves at varying speed between node pair, are comparatively much higher than no car movement. The similar trend is also shown in Fig. 7 and it shows that the average RSSI value when car moves at varying speed and no car movement between the node pair. Conversely, the standard deviation in RSSI decreases when car speed increases, as shown in Fig. 8. The reason of this high deviation at slow speed is that the car obstructs the signal for longer time compared to faster speed and results in a high deviation.

\begin{figure}
\center
\includegraphics[height=4.2cm]{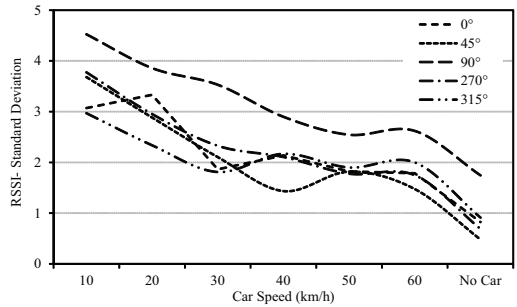}
\caption{RSSI standard deviation vs. varying car speed for different antenna orientation angles}
\label{Fig_8}
\end{figure}

The RSSI variation envelope (maximum, average and minimum values) for 10, 30 and 40km/h car speed versus varying elevation angle of transmitter antenna is shown in Fig. 9 (a), (b) and (c), respectively. The RSSI envelope size of different speeds at different elevation angles is almost similar for different angles. However, some irregularities have been observed in RSSI values at 45  in Fig. 9 (c) and 90  in Fig. 9. The RSSI deviation is less at 45  when car moves at 40km/h and the RSSI has very high deviation in maximum or high RSSI direction with respect to the average RSSI value at 90  for all speeds. The similar trend is shown and summarized in Fig. 10.

\begin{figure}
\center
\includegraphics[height=10.2cm]{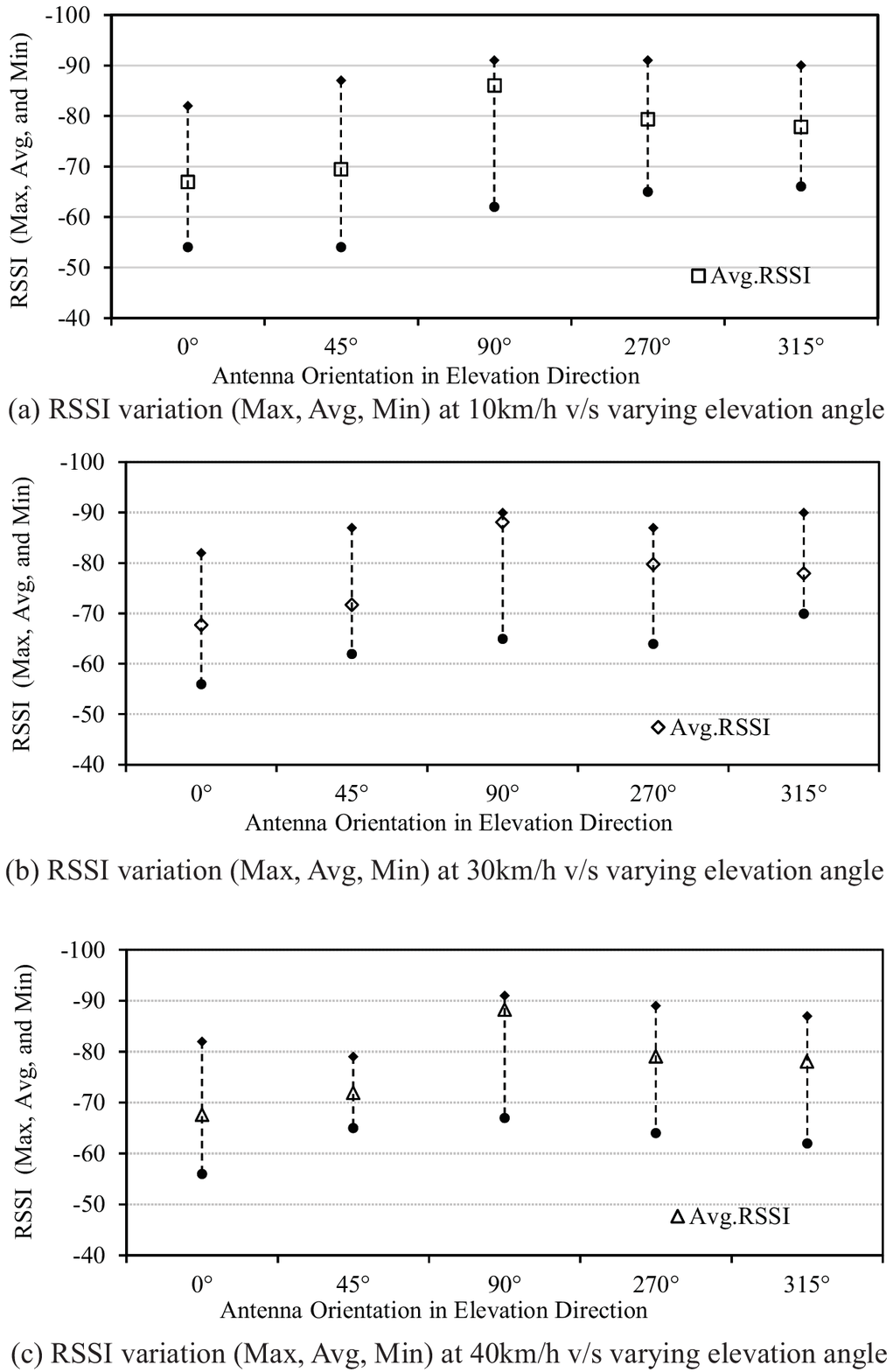}
\caption{RSSI variation vs. varying elevation angles at (a) 10km/h, (b) 30km/h and (c) 40km/h car speed.}
\label{Fig_9}
\end{figure}

\begin{figure}
\center
\includegraphics[height=4.2cm]{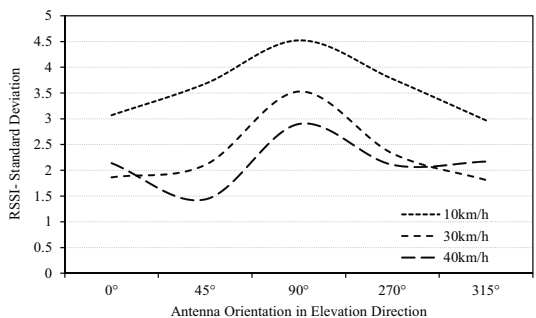}
\caption{RSSI Standard Deviation vs. varying elevation angles for different car speed.}
\label{Fig_10}
\end{figure}

It has been observed from the experimental results that antenna elevation in presence of the ground effect has some irregularities in the RSSI fluctuations. Hence, to detect the fast moving object by observing the RSSI value, an efficient algorithm is needed.

\section{Conclusion}\label{sec:conclusion}

In this paper, we experimentally analyzed the effect of fast moving vehicle on the RSSI in presence of the ground effect and antenna orientation in elevation direction. It has been observed that the slow speed causes high deviation in RSSI compared to the faster speed. Some irregularities in the RSSI fluctuations are also observed when antenna is elevated at 45  and 90 . 
In future, this work will be extend to propose the fast moving object sensing algorithm for SL-WSN that takes into account the antenna orientation as well as ground effect. Algorithm must incorporate the learning phase beforehand to precisely detect the fast moving object with very low false count.

\subsubsection*{Acknowledgments.} This work was partly supported by Keio University $21^{st}$ Century Center of Excellence Program on ``Optical and Electronic Device Technology for Access Network'' and Fujitsu Laboratories

\end{document}